\documentclass[aps,twocolumn]{revtex4-1}
\usepackage{amsmath}
\usepackage{graphicx}
\usepackage{color}

\begin{document}

\title{Superconducting state generated dynamically from distant pair source
and drain}
\author{E. S. Ma}
\author{Z. Song}
\email{songtc@nankai.edu.cn}
\affiliation{School of Physics, Nankai University, Tianjin 300071, China}
\begin{abstract}
It has been well established that the origin of $p$-wave superconductivity
is the balance between pair creation and annihilation, described by the
spin-less fermionic Kitaev model. In this work, we study the dynamics of a
composite system where the pair source and drain are spatially separated by
a long distance. We show that this non-Hermitian system possesses a
high-order exceptional point (EP) when only a source or drain is considered.
The EP dynamics provide a clear picture: A pair source can fully fill the
system with pairs, while a drain can completely empty the system. When the
two coexist simultaneously, the dynamics depend on the distance and the
relative phase between the pair creation and annihilation terms. Analytical
analysis and numerical simulation results show that the superconducting
state can be dynamically established at the resonant pair source and drain:
from an initial empty state to a stationary state with the maximal pair
order parameter. It provides an alternative way of understanding the
mechanism of the nonequilibrium superconducting state.
\end{abstract}

\maketitle

\section{Introduction}

Motivated by recent advances in experimental capability \cite%
{kinoshita2006quantum,trotzky2012probing,gring2012relaxation,schreiber2015observation,smith2016many,kaufman2016quantum}%
, the nonequilibrium dynamics of quantum many-body systems has emerged as a
fundamental and attractive topic in condensed-matter physics. As one of the
potential applications, nonequilibrium many-body dynamics provide an
alternative way to access a new exotic quantum state with energy far from
the ground state \cite%
{Choi,Else,Khemani,Lindner,Kaneko,Tindall,YXMPRA,zhang2022steady}. Unlike
traditional protocols based on the cooling mechanism, quench dynamics have a
wide range of potential applications since they provide many ways to take a
system out of equilibrium, such as applying a driving field or pumping
energy or particles in the system through external reservoirs {\cite{QD1,
QD2,QD3}}. This makes it possible to design interacting many-body systems to
prepare some desirable many-body quantum states by a quenching process. Much
effort \cite{kitamura2016eta,kaneko2019photoinduced,tindall2019heating,
fujiuchi2019photoinduced,peronaci2020enhancement,fujiuchi2020superconductivity, ejima2020photoinduced,li2020eta,kaneko2020charge,zhang2021eta,zhang2020dynamical,yang2022dynamic,zhang2022steady}
has been devoted to developing various nonequilibrium protocols for the
generation of the $\eta $-pairing-like state \cite{yang1989eta} in the
Hubbard model.

In parallel, the Kitaev model is a lattice model of a $p$-wave
superconducting wire, which realize Majorana zero modes at the ends of the
chain \cite{Kitaev}. This has been demonstrated by unpaired Majorana modes
exponentially localized at the ends of open Kitaev chains \cite%
{Sarma,Stern,Alicea}. The main feature of this model originates from the
pairing term, which violates the conservation of the fermion number but
preserves its parity, leading to the superconducting phase. The amplitudes
for pair creation and annihilation play an important role in the existence
of the gapped superconducting phase. Compared to the Hubbard model, the
Kitaev model has an advantage for the task since it is exactly solvable. In
recent work on a simple 1D Kitaev model \cite{shi2022dynamic,YXMPRA}, it has
been shown that a nonequilibrium superconducting state can be obtained
through time evolution from an initially prepared vacuum state, providing an
alternative approach to dynamically generate a superconducting state from an
easily prepared trivial state.

Due to the translational symmetry of such a system, every pair term
contributes equally to the formation of the superconducting state. A
question arises as to what happens if there is only a single pair term in
the Hamiltonian. In this work, we study the dynamics of a composite system
where the pair source and drain are considered individually or spatially
separated by a long distance. In the framework of quantum mechanics, it
corresponds to a non-Hermitian Hamiltonian \cite{Bender,Bender1}. Many
contributions have been devoted to non-Hermitian Kitaev models \cite%
{Law,Tong,Yuce,You,Klett,Menke} and Ising models \cite%
{LCPRB,LC2014conventional,LC2015finite,LC2016chern,ZXZ} within the
pseudo-Hermitian framework. We show that this non-Hermitian system possesses
a high-order exceptional point (EP) when only a source or drain is
considered. It admits peculiar dynamics: the final state is a particular
eigenstate, coalescing state {\cite{Berry2004,Heiss2012,Miri2019,Zhang2020}}%
. The EP dynamics provide a clear physical picture: A pair source can
eventially fully fill the system with pairs, while a drain can completely
empty the system. However, both final states are trivial. When the two
coexist simultaneously, the dynamics depend on the distance and the relative
phase between the pair creation and annihilation terms. Analytical analysis
and numerical simulation results for a finite system show that a perfect
superconducting state can be dynamically established at the resonant pair
source and drain, i.e., an initial empty state evolves to a stationary
state, which is a perfect superconducting state with the maximal pair order
parameter. We consider a composite system with nonhomogenous pair terms in
the present work in comparison to previous works.\ This provides an
alternative mechanism for forming nonequilibrium superconducting state.

This paper is organized as follows. In Section \ref{Hamiltonian and order
parameter}, we describe the model Hamiltonian and introduce the order
parameter. In Section \ref{Perfect superconducting state}, based on the
exact solutions of a toy model, we demonstrate that its ground state has the
maximal order parameter. In Section \ref{High-order EP and dynamics}, we
study the dynamics in the non-Hermitian system, where only pair creation or
annihilation terms are considered. In Section \ref{Single source or drain},
we investigate the dynamics driven by a single source or drain.\ In Section %
\ref{Distant source and drain}, we investigate the dynamics driven by
spatially separated source and drain at resonance. Finally, we give a
summary and discussion in Section \ref{sec_summary}.

\begin{figure}[tbh]
\centering \includegraphics[width=0.4\textwidth]{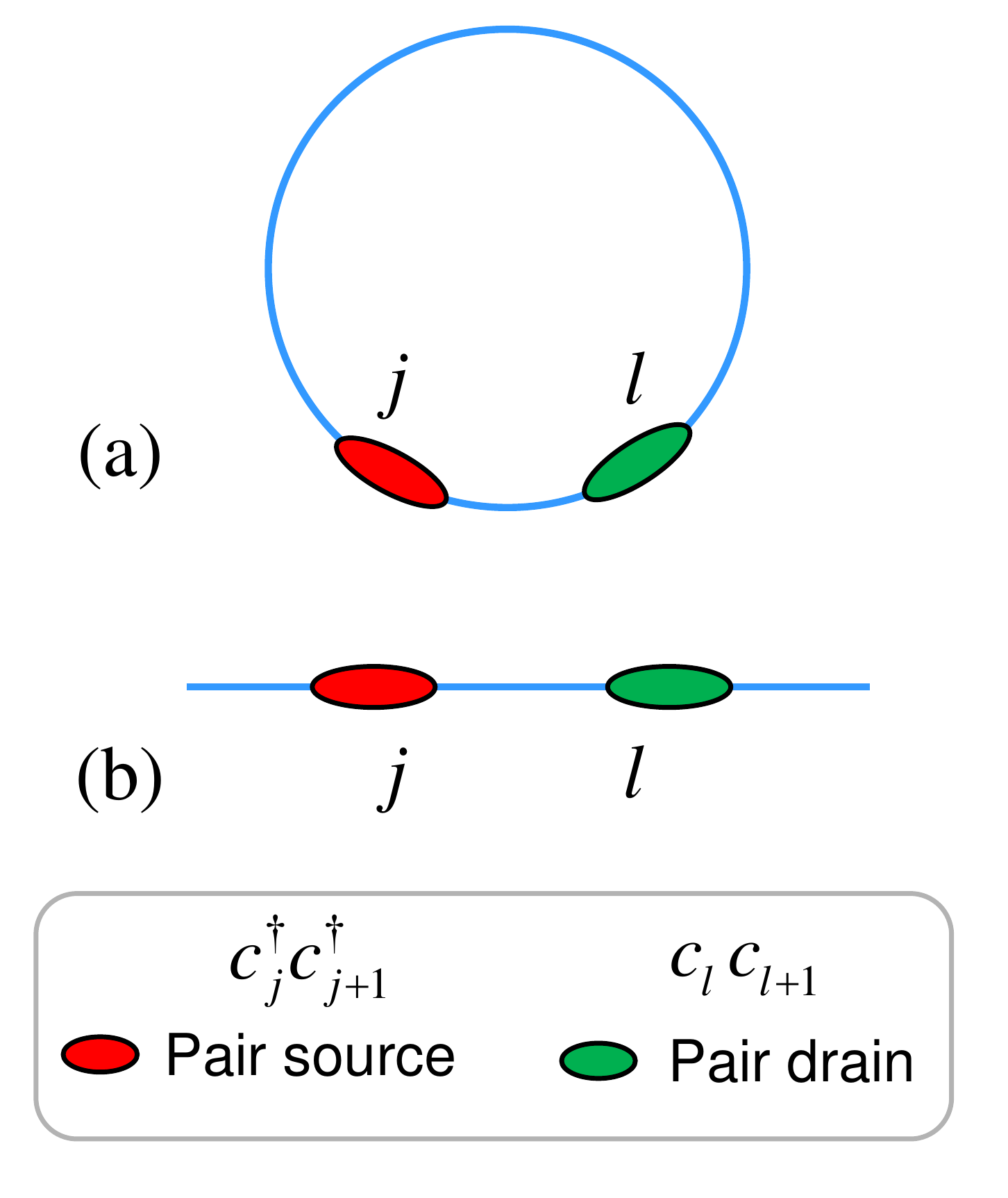}
\caption{Schematic of the 1D Kitaev models for spinless fermions with pair
terms across two adjacent sites, red and green dimers indicating specially
separated pair creation (source) and annihilation (drain), respectively. Two
resonant pair terms are embedded in a (a) ring lattice and (b) open chain.
The goal of this work is to investigate the effect of the setup on the
nonequilibrium state after sufficiently long time.}
\label{fig1}
\end{figure}

\section{Hamiltonian and order parameter}

\label{Hamiltonian and order parameter}

We start with a generalized Hamiltonian%
\begin{eqnarray}
H &=&H_{\mathrm{T}}+H_{\mathrm{P}}, \\
H_{\mathrm{T}} &=&-iT\sum\limits_{j=1}^{N}c_{j}^{\dag }c_{j+1}+\mathrm{H.c.}%
+\mu \sum\limits_{j=1}^{N}n_{j},  \label{HT} \\
H_{\mathrm{P}} &=&\sum\limits_{j=1,r>0}^{N}\left( f_{j}^{r}c_{j}^{\dag
}c_{j+r}^{\dag }+g_{j}^{r}c_{j+r}c_{j}\right) ,
\end{eqnarray}%
where the distribution functions $f_{j}^{r}$\ and $g_{j}^{r}$ determine the
location of creation and annihilation of a pair with size $r$, respectively.
In the case with $f_{j}^{r}=g_{j}^{r}=\Delta \delta _{r,1}$ and $%
iT\rightarrow T$, the model as a paradigm of $p$ wave topological
superconductivity has been well studied. Here, we take the imaginary hopping
strength for the sake of convenience in the following discussion, and $%
iT\rightarrow T$\ can be realized by local gauge transformation. However,
the origin of the phase in the hopping cannot be considered as the magnetic
flux since it still takes effect on the eigenstates even if the open
boundary condition is taken due to the existence of the pair term. One of
the possible origins is the phase gradient on the pair term \cite%
{seradjeh2011unpaired,romito2012manipulating,rontynen2014tuning,dmytruk2019majorana,takasan2022supercurrent}%
.

The main purposes of the following discussion are (i) to present a Hermitian
system with a superconducting ground state, which possesses the maximal
BCS-pair order parameter, and (ii) to provide a scheme to achieve such a
state dynamically by a non-Hermitian system with local pair source and
drain. To this end, we will consider several types of functions $\left\{
f_{j}^{r},g_{j}^{r}\right\} $, which correspond to different types of
pairing processes, including local and extensive sized pairing. In the case
with spatially separated creation and annihilation of pairs, the
non-Hermitian term is naturally involved, and then some particular dynamic
behaviors may emerge that never appear in a Hermitian system. One of them is
the EP dynamics, which provides a mechanism for a relaxation process in the
framework of quantum mechanics. In fact, in the case with $%
f_{j}^{r}=g_{j}^{r}=\Delta \delta _{r,1}$, the model has been studied
systematically \cite{Off-diagonal2022ma}. It has been
shown that the ground states near the critical point $\mu =0$\ possess ODLRO
in association with the maximum of BCS-pair order parameter.

In the framework of the Kitaev model, the pair number is not suitable for
characterizing a superconducting state, since the fully filled pair state $%
\prod_{k>0}c_{-k}^{\dag }c_{k}^{\dag }\left\vert 0\right\rangle $ $%
=e^{i\theta }\prod_{l=1}^{N}c_{l}^{\dag }\left\vert 0\right\rangle $ is an
insulating state. To quantitatively characterize the superconductivity of a
given state $\left\vert \psi \right\rangle $, we introduce the operator%
\begin{equation}
\mathcal{O}=\frac{2}{N}\sum_{k>0}|\left\langle \psi \right\vert
c_{k}c_{-k}\left\vert \psi \right\rangle |.
\end{equation}%
Obviously, for a given state $\left\vert \psi \right\rangle $, quantity $%
|\left\langle \psi \right\vert c_{k}c_{-k}\left\vert \psi \right\rangle |$ $%
=|\left\langle \psi \right\vert c_{-k}^{\dag }c_{k}^{\dag }\left\vert \psi
\right\rangle |$ measures the rate of transition for a pair at $k$ channel
and the population of pairs. Then, $\mathcal{O}$ is defined by the average
magnitude over all channels. In general, nonzero $\mathcal{O}$ means that
state $\left\vert \psi \right\rangle $ is a superconducting state.

\section{Perfect superconducting state}

\label{Perfect superconducting state}

In this section, we consider a specific distribution function

\begin{equation}
D(r)=-\frac{2i}{\pi }\frac{\Delta }{r}\delta _{r,\mathrm{odd}},
\end{equation}%
which is the key of the following toy model. Such a toy model allows us to
obtain an exact solution, which is ultimately related to the main goal of
this work. To this end, we introduce a set of pseudo spin operators%
\begin{eqnarray}
s^{-} &=&\left( s^{+}\right) ^{\dag }=\sum\limits_{k>0}c_{k}c_{-k}, \\
s^{z} &=&\frac{1}{2}\sum\limits_{k>0}\left( c_{k}^{\dag }c_{k}+c_{-k}^{\dag
}c_{-k}-1\right) ,
\end{eqnarray}%
which obey the SU(2) commutation relation, $\left[ s^{+},s^{-}\right]
=2s^{z} $. Particularly, in real space, the spin operator has the form%
\begin{equation}
s^{-}=\frac{2i}{\pi }\sum\limits_{j}\sum\limits_{\mathrm{odd}\text{ }r}\frac{%
1}{r}c_{j}c_{j+r},
\end{equation}%
in the thermodynamic limit $N\longrightarrow \infty $. Then, taking $%
f_{j}^{r}=\left( g_{j}^{r}\right) ^{\ast }=D^{\ast }(r)$, we have 
\begin{equation}
H_{\mathrm{P}}=\frac{2\Delta i}{\pi }\sum\limits_{j}\sum\limits_{\mathrm{odd}%
\text{ }r}\frac{1}{r}\left( c_{j}^{\dag }c_{j+r}^{\dag }-c_{j+r}c_{j}\right)
,
\end{equation}%
and the Hermitian extended Kitaev Hamiltonian in the form%
\begin{eqnarray}
H_{\mathrm{ext}} &=&2T\sum\limits_{k>0}\sin k(c_{k}^{\dag
}c_{k}-c_{-k}^{\dag }c_{-k})+2\mu s^{z}  \notag \\
&&+\Delta \left( s^{+}+s^{-}\right) .
\end{eqnarray}

Here, we neglect a constant $\mu N/2$\ for the sake of simplicity. In the
BCS-pair invariant subspace spanned by the pair states 
\begin{equation}
\left\vert \Psi _{\left\{ k\right\} }\right\rangle
=\prod_{\{k\}}c_{-k}^{\dag }c_{k}^{\dag }\left\vert 0\right\rangle ,
\end{equation}%
where $\left\{ k\right\} $\ denotes the $2^{N/2}$\ dimensional set of
configuration of the BCS-pair filling, we always have $(c_{k}^{\dag
}c_{k}-c_{-k}^{\dag }c_{-k})\left\vert \Psi _{\left\{ k\right\}
}\right\rangle $ $=0\ $and $(c_{k}^{\dag }c_{k}-c_{-k}^{\dag
}c_{-k})\left\vert 0\right\rangle $ $=0$. This indicates that all $2^{N/2}$\
states $\left\{ \left\vert \Psi _{\left\{ k\right\} }\right\rangle \right\} $%
\ are zero-energy eigenstates of $H_{\mathrm{T}}$, i.e., the first term of $%
H_{\mathrm{ext}}$. Then, we have the equivalent Hamiltonian of $H_{\mathrm{%
ext}}$ in the subspace%
\begin{equation}
\mathcal{H}_{\mathrm{ext}}=\mathbf{B}\cdot \mathbf{s},
\end{equation}%
where the magnetic field $\mathbf{B=}(2\Delta ,0,2\mu )$. Obviously, both $%
\mathbf{s}^{2}$\ and $s^{z}$\ are commutative to $H_{\mathrm{ext}}$. Based
on this fact, one can further construct multi-invariant subspaces by the
common eigenstates of $\mathbf{s}^{2}$\ and $s^{z}$. We are interested in a
set $\left\{ \left\vert \psi _{n}\right\rangle \right\} $ with $n\in \left[
0,N/2\right] $ 
\begin{equation}
\left\vert \psi _{n}\right\rangle =\frac{1}{\Omega _{n}}(s^{+})^{n}\left%
\vert 0\right\rangle ,
\end{equation}%
which obeys $\mathbf{s}^{2}\left\vert \psi _{n}\right\rangle =$ $N\left(
N/4+1\right) /4\left\vert \psi _{n}\right\rangle $\ and $s^{z}\left\vert
\psi _{n}\right\rangle =$ $\left( n-N/4\right) \left\vert \psi
_{n}\right\rangle $, with the normalization factor $\Omega _{n}=$ $\left(
n!\right) \sqrt{C_{N/2}^{n}}$. The eigenstates of $H_{\mathrm{ext}}$ are
actually the eigenstates of the spin operator%
\begin{equation}
\frac{\mathbf{B}}{\left\vert \mathbf{B}\right\vert }\cdot \mathbf{s=}\frac{1%
}{\sqrt{\Delta ^{2}+\mu ^{2}}}(\Delta s^{x}+\mu s^{z}),
\end{equation}%
and be expressed in the form%
\begin{equation}
\left\vert \Phi _{l}(\mu )\right\rangle =\sum_{n}d_{n}^{l}(\mu )\left\vert
\psi _{n}\right\rangle ,
\end{equation}%
satisfying the equation%
\begin{equation}
H_{\mathrm{ext}}\left\vert \Phi _{l}(\mu )\right\rangle =2\sqrt{\Delta
^{2}+\mu ^{2}}\left( l-\frac{N}{4}\right) \left\vert \Phi _{l}(\mu
)\right\rangle .
\end{equation}%
The coefficient $d_{n}^{l}(\mu )$ can be obtained exactly, but here, we only
list two of them explicitly%
\begin{eqnarray}
&&d_{n}^{0}(\mu )=\sqrt{C_{N/2}^{n}}\left( -1\right) ^{n}\cos ^{n}\frac{%
\delta }{2}\sin ^{\left( N/2-n\right) }\frac{\delta }{2}, \\
&&d_{n}^{N/2}(\mu )=\sqrt{C_{N/2}^{n}}\sin ^{n}\frac{\delta }{2}\cos
^{\left( N/2-n\right) }\frac{\delta }{2},
\end{eqnarray}%
which lead to

\begin{eqnarray}
&&\left\vert \Phi _{0}(\mu )\right\rangle =\prod\limits_{k>0}\left( \sin 
\frac{\delta }{2}-\cos \frac{\delta }{2}c_{-k}^{\dag }c_{k}^{\dag }\right)
\left\vert 0\right\rangle , \\
&&\left\vert \Phi _{N/2}(\mu )\right\rangle =\prod\limits_{k>0}\left( \cos 
\frac{\delta }{2}+\sin \frac{\delta }{2}c_{-k}^{\dag }c_{k}^{\dag }\right)
\left\vert 0\right\rangle ,
\end{eqnarray}%
with $\tan \delta =-\Delta /\mu $. We note that the two above eigenstates
reduce to $\prod_{k>0}(1\mp $ $c_{-k}^{\dag }c_{k}^{\dag })$ $/\sqrt{2}%
\left\vert 0\right\rangle _{k}\left\vert 0\right\rangle _{-k}$,\ which
supports that the corresponding order parameter reaches the maximum $0.5$\
when taking the chemical potential $\mu =0$. We refer to such states as
perfect superconducting states.

It is clear that these two states are not unique perfect superconducting
states. In fact, state $\prod_{k>0}(1+$ $e^{i\gamma _{k}}c_{-k}^{\dag
}c_{k}^{\dag })$ $/\sqrt{2}\left\vert 0\right\rangle $\ has the same feature
for any distribution of $\left\{ \gamma _{k}\right\} $. This inspires us to
find another way to prepare a superconducting state. Now we consider the
dynamic generation of such\ states in the case with zero $\mu $. For the
initial state $\left\vert \psi \left( 0\right) \right\rangle =\left\vert
\psi _{0}\right\rangle $, which is actually an empty state, the time
evolution under a quench Hamiltonian 
\begin{equation}
H_{\mathrm{quen}}=\Delta (s^{-}+s^{+}),  \label{H_ext real}
\end{equation}%
can be expressed as%
\begin{equation}
\left\vert \psi \left( t\right) \right\rangle =\exp \left[ -i\Delta
(s^{-}+s^{+})t\right] \left\vert \psi \left( 0\right) \right\rangle ,
\end{equation}%
which is essentially a rotation around the $x$-axis. Obviously $\left\vert
\psi \left( t\right) \right\rangle $\ can be easily obtained by local
rotation for each $k$, 
\begin{equation}
\left\vert \psi \left( t\right) \right\rangle =\prod\limits_{k>0}\left( \cos
\theta -i\sin \theta c_{-k}^{\dag }c_{k}^{\dag }\right) \left\vert
0\right\rangle ,
\end{equation}%
where $\theta $\ is a function of time $\theta (t)=\Delta t$. Direct
derivation shows that the order parameter is a periodic function of time%
\begin{equation}
\mathcal{O}(t)=\frac{1}{2}\left\vert \sin (2\Delta t)\right\vert ,
\end{equation}%
which reaches $0.5$\ at instants $t=(m+1/2)\pi /(2\Delta )$ with integer $m$%
. This indicates that a perfect superconducting state can be established via
a dynamic process. The physics seems to be clear that the oscillating $%
\mathcal{O}(t)$\ is a resultant effect of both pair creation and
annihilation terms.

\section{High-order EP and dynamics}

\label{High-order EP and dynamics}

Now, we consider a question of what happens if only the pair annihilation
(creation) terms are taken. It is a first step to investigate the effect of
spatially separated source and drain. Naturally, a non-Hermitian Hamiltonian
by taking $f_{j}^{r}=0$ but $g_{j}^{r}=D(r)$ is involved, i.e.,

\begin{equation}
H_{\mathrm{P}}=-\frac{2i\Delta }{\pi }\sum\limits_{j=1}^{N}\sum\limits_{%
\mathrm{odd}\text{ }r}\frac{1}{r}c_{j+r}c_{j}.
\end{equation}%
The equivalent Hamiltonian in the invariant subspace spanned by the set of
states $\left\{ \left\vert \psi _{n}\right\rangle \right\} $ becomes%
\begin{equation}
\mathcal{H}_{\mathrm{ext}}=\Delta s^{-}.
\end{equation}%
The dynamics of $\mathcal{H}_{\mathrm{ext}}$\ are slightly little special
and can be captured from the matrix representation of Hamiltonian $\mathcal{H%
}_{\mathrm{ext}}$. It is an $\left( N/2+1\right) \times \left( N/2+1\right) $
matrix $M$, with nonzero matrix elements

\begin{equation}
\left( M\right) _{N/2-n,N/2+1-n}=\Delta \sqrt{n\left( N/2-n+1\right) }\text{,%
}
\end{equation}%
with $n=\left[ 0,N/2-1\right] $. Note that $M$ is a nilpotent matrix, i.e. $%
M^{N/2+1}=0$,\ or an $\left( N/2+1\right) $-order Jordan block. The dynamics
for any state\ in this subspace $\left\{ \left\vert \psi _{n}\right\rangle
\right\} $ is governed by the time evolution operator%
\begin{equation}
U(t)=e^{-iMt}=\sum_{l=0}^{N/2}\frac{1}{l!}\left( -iMt\right) ^{l}.
\end{equation}%
Then for the initial state $\left\vert \psi \left( 0\right) \right\rangle
=\left\vert \psi _{N/2}\right\rangle $, we have the normalized evolved state%
\begin{equation}
\left\vert \psi \left( t\right) \right\rangle =\prod_{k>0}\frac{-it\Delta
+c_{-k}^{\dag }c_{k}^{\dag }}{\sqrt{1+\Delta ^{2}t^{2}}}\left\vert
0\right\rangle ,
\end{equation}%
which turns to the coalescing state, i.e., $\left\vert \psi \left( \infty
\right) \right\rangle \longrightarrow \left\vert \psi _{0}\right\rangle $.
Obviously, $\left\vert \psi \left( t\right) \right\rangle $\ has maximal $%
\mathcal{O}$ at instant $1/\Delta $. The above analysis is still true when
we take $f_{j}^{r}=D^{\ast }(r)$ but $g_{j}^{r}=0$\ and $\left\vert \psi
\left( 0\right) \right\rangle =\left\vert \psi _{0}\right\rangle $, which
corresponds to a time reversal process. As expected, the physical picture is
clear that the pair term takes the role of not only pair generation but also
reduction. Intuitively, a local pair term should have a similar effect. This
is the aim of the next section.

\section{Single source or drain}

\label{Single source or drain}

The results obtained in the last section are exact and explicit due to the
translational symmetry of the model. In this section, we will show that a
similar result can be obtained approximately when only a single pair term is
considered, i.e., $f_{j}^{r}=0$ but $g_{j}^{r}=\delta _{j,j_{0}}D(r)$, or
vice versa. First, states $\left\{ \left\vert \psi _{n}\right\rangle
\right\} $\ have translational symmetry with zero momentum. This originates
from the translational symmetry of the system, i. e., $\left[ H_{\mathrm{T}},%
\mathcal{T}_{1}\right] =0$, where operator $\mathcal{T}_{1}$\ is defined by $%
\mathcal{T}_{1}c_{j}\mathcal{T}_{1}^{-1}=c_{j+1}$. Obviously, we have $%
\mathcal{T}_{1}\left\vert \psi _{0}\right\rangle =\left\vert \psi
_{0}\right\rangle $, which results in%
\begin{equation}
\mathcal{T}_{1}\left\vert \psi _{n}\right\rangle =\alpha \left\vert \psi
_{n}\right\rangle ,
\end{equation}%
with $\left\vert \alpha \right\vert =1$ because $\mathcal{T}_{1}^{-1}s^{\pm }%
\mathcal{T}_{1}=s^{\pm }$. Then, we have if $N\longrightarrow \infty $

\begin{equation}
\left\langle \psi _{n}\right\vert \sum\limits_{\mathrm{odd}\text{ }r}\frac{1%
}{r}c_{j+r}c_{j}\left\vert \psi _{m}\right\rangle =\frac{\pi i}{2N}\sqrt{%
m\left( N/2-m+1\right) }\delta _{n,m-1},
\end{equation}%
based on the relation%
\begin{equation}
\left\langle \psi _{n}\right\vert c_{j+r}c_{j}\left\vert \psi
_{m}\right\rangle =\left\langle \psi _{n}\right\vert
c_{j+r+1}c_{j+1}\left\vert \psi _{m}\right\rangle .
\end{equation}%
Obviously, the perturbation matrix is still in ($N+1$)-order Jordan block
form. The time evolution under such a system should obey the EP dynamics.

Furthermore, a Jordan block matrix does not restrict the values of the
nonzero matrix elements. Then the relation 
\begin{equation}
\left\langle \psi _{n}\right\vert c_{j+1}c_{j}\left\vert \psi
_{m}\right\rangle \propto \delta _{m,n+1},
\end{equation}%
may also result in EP dynamics, based on the following analysis. Actually,
considering a more generalized form of $H_{\mathrm{P}}=\sum_{i,j}\lambda
_{ij}c_{i}c_{j}$ with arbitrary factor $\left\{ \lambda _{ij}\right\} $,
states $\left\vert \psi _{0}\right\rangle $\ and $\left\vert \psi
_{N}\right\rangle $\ are two degenerate states of the Hermitian Hamiltonian $%
H_{\mathrm{T}}$, and we always have%
\begin{equation}
\mathcal{H}\left\vert \psi _{0}\right\rangle =0,\mathcal{H}^{\dag
}\left\vert \psi _{N/2}\right\rangle =0,
\end{equation}%
due to the facts 
\begin{equation}
c_{j+r}c_{j}\left\vert \psi _{0}\right\rangle =0,\left( c_{j+r}c_{j}\right)
^{\dag }\left\vert \psi _{N/2}\right\rangle =0.
\end{equation}%
This means that two states $\left\vert \psi _{0}\right\rangle $\ and that $%
\left\vert \psi _{N/2}\right\rangle $\ are mutually biorthogonal conjugates
and $\langle \psi _{0}\left\vert \psi _{N/2}\right\rangle $ is their
biorthogonal norm. Importantly, the vanishing norm $\langle \psi
_{0}\left\vert \psi _{N/2}\right\rangle =0$ indicates that state $\left\vert
\psi _{0}\right\rangle $($\left\vert \psi _{N/2}\right\rangle $) is the
coalescing state of $H$($H^{\dag }$) or Hamiltonians $H$\ and $H^{\dag }$\
obtain an EP. From the perspective of dynamics, we have%
\begin{equation}
e^{-iHt}\left\vert \psi _{N/2}\right\rangle \longrightarrow \left\vert \psi
_{0}\right\rangle ,e^{-iH^{\dag }t}\left\vert \psi _{0}\right\rangle
\longrightarrow \left\vert \psi _{N/2}\right\rangle ,
\end{equation}%
for a sufficiently long time $t$. Although both states $\left\vert \psi
_{0}\right\rangle $\ and $\left\vert \psi _{N/2}\right\rangle $\ are trivial
states, $e^{-iHt}\left\vert \psi _{N/2}\right\rangle \ $and $e^{-iH^{\dag
}t}\left\vert \psi _{0}\right\rangle $ may have pair currents at finite $t$. 
\begin{figure*}[tbh]
\centering
\includegraphics[width=1\textwidth]{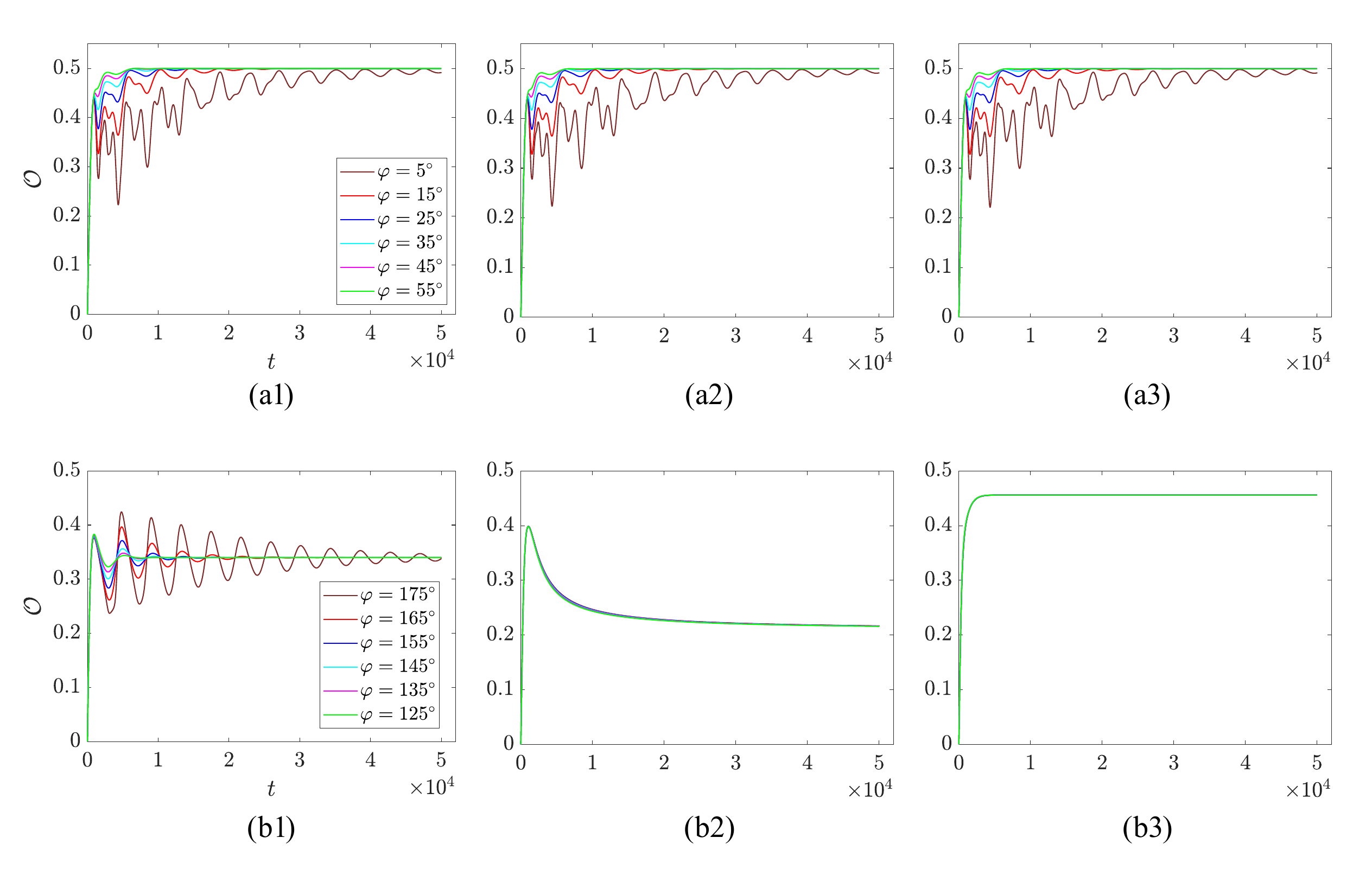}
\caption{Plots of the time evolution of $\mathcal{O}$ in Eq. (\protect\ref%
{Ot}) for several {representative }$\protect\varphi $ under the Hamiltonian
with $H_{\mathrm{T}}$ in Eq. (\protect\ref{HT}) and $H_{\mathrm{quen}}$ in
Eq. (\protect\ref{Hq}) on the lattice, {which is schematically illustrated
in Fig. 1 (a) and (b), respectively.} The initial state is the vacuum state
and the parameters are $N=9$, $T=1$, $\protect\mu =0$ and $\Delta =0.005$.
(a1), (a2) and (a3) are the situations for a ring lattice with $N_{0}=$ $3$, 
$4$ and $5$, respectively. (b1), (b2) and (b3) are the situations for an
open chain with $N_{0}=$ $3$, $5$ and $8$, respectively. {We find that for
all cases, $\mathcal{O}(t)$\ tends to } stabilize at {$\mathcal{O}(\infty )$}
after a sufficiently long time. The results show that the stable final state
has maximal $\mathcal{O}$ for different $\protect\varphi $. {$\mathcal{O}%
(\infty )$}\ is not sensitive to $N_{0}$ for the ring system, and $\protect%
\varphi $ affects the rate of convergence of the evolved state. For the
chain system, {$\mathcal{O}(\infty )$} reaches the maximum when the pair
source and drain are located at the ends of the chain.\ }
\label{fig2}
\end{figure*}

\section{Resonant distant source and drain}

\label{Distant source and drain}

The above result at least indicates that the local pair terms can be
regarded as particle sources or drains, which can fully fill the empty state
($\left\vert \psi _{0}\right\rangle \longrightarrow \left\vert \psi
_{N/2}\right\rangle $)\ or empty the fully filled state ($\left\vert \psi
_{N/2}\right\rangle \longrightarrow \left\vert \psi _{0}\right\rangle $).\
This inspires us to investigate the dynamics with balanced local pair terms
that are spatially separated by a distance. To this end, we consider the
pair term in the resonant form

\begin{equation}
f_{j}^{r}=e^{i\varphi }\delta _{j,1}D^{\ast }(r),g_{j}^{r}=\delta
_{j,N_{0}}D(r),
\end{equation}%
or explicitly%
\begin{equation}
H_{\mathrm{P}}=\frac{2i\Delta }{\pi }\sum\limits_{\mathrm{odd},r}\frac{1}{r}%
\left( e^{i\varphi }c_{1}^{\dag }c_{1+r}^{\dag }-c_{N_{0}+r}c_{N_{0}}\right)
,
\end{equation}%
which acts as separated local pair sources and drains. Here, as the resonant
condition, the amplitudes of the pair annihilation and creation terms are
the same, while\ there is a phase difference $\varphi $ between them, which
is crucial for the dynamics, as shown in the following. In the small $\Delta 
$\ limit, based on the perturbation method, the matrix representation of
Hamiltonian $\mathcal{H}_{\mathrm{ext}}$ in the subspace spanned by the set
of states $\left\{ e^{in\varphi /2}\left\vert \psi _{n}\right\rangle
\right\} $ is an $\left( N/2+1\right) \times \left( N/2+1\right) $ matrix $%
\mathcal{H}_{\mathrm{ext}}$ with nonzero matrix elements

\begin{eqnarray}
\left( \mathcal{H}_{\mathrm{ext}}\right) _{N/2+1-n,N/2-n} &=&e^{i\varphi /2}%
\frac{\Delta }{N}\sqrt{n\left( N/2+1-n\right) }  \notag \\
&=&\left( \mathcal{H}_{\mathrm{ext}}\right) _{N/2-n,N/2+1-n},
\label{H_ext complex}
\end{eqnarray}%
with $n=\left[ 0,N/2-1\right] $. We note that matrix $\mathcal{H}_{\mathrm{%
ext}}$ is the same as that in (\ref{H_ext real}) but with a complex strength
constant. The corresponding eigenenergy is complex%
\begin{equation}
E_{n}=\frac{\Delta e^{i\varphi /2}}{N}(2n-\frac{N}{2}),
\end{equation}%
with $n=\left[ 0,N/2\right] $, and its imaginary part is $\mathrm{Im}\left(
E_{n}\right) =$ $\Delta \left( 2n/N-1/2\right) \sin \frac{\varphi }{2}$.
Unlike a Hermitian system, the imaginary part of the eigenvalue can amplify
or reduce\ the corresponding amplitude of the wave function in the dynamic
process. For the given initial state $\left\vert \psi \left( 0\right)
\right\rangle =\left\vert \psi _{0}\right\rangle $ when the evolution time
is sufficiently long, the final state is the eigenstate of $\mathcal{H}_{%
\mathrm{ext}}$ with the maximum imaginary part. The corresponding
approximate eigenstate is 
\begin{equation}
\left\vert \psi \left( \infty \right) \right\rangle =\prod\limits_{k>0}\frac{%
\sigma e^{-i\varphi /2}+c_{-k}^{\dag }c_{k}^{\dag }}{\sqrt{2}}\left\vert
0\right\rangle ,  \label{final state}
\end{equation}
where $\sigma =\mathrm{sgn}(\sin \frac{\varphi }{2})$. Obviously, $%
\left\vert \psi \left( \infty \right) \right\rangle $\ is a perfect
superconducting state. When the off-resonant case is considered, the
expression is the same as $\left\vert \psi \left( \infty \right)
\right\rangle $, but an imaginary part should be added in $\varphi $, which
will reduce the order parameter from $0.5$.

Now we consider a more practical case with%
\begin{equation}
\mathcal{H}_{\mathrm{quen}}=\Delta \left( e^{i\varphi }c_{2}^{\dag
}c_{1}^{\dag }+c_{N_{0}}c_{N_{0}+1}\right) ,  \label{Hq}
\end{equation}%
where the pairing terms reduce to the simplest case. In addition, we also
consider the case with an open boundary condition, which is closer to the
real sample in the experiment. The physical intuition for this setup is
simple. Term $c_{1}^{\dag }c_{2}^{\dag }$\ acts as a source of pair at one
end of the chain, while $c_{N-1}c_{N}$ takes the role of drain at the other
end. According to the analysis of the pair term $\sum_{i,j}\lambda
_{ij}c_{i}c_{j}$\ in the last section, it is expected that the nearest
neighboring pair terms share a similar feature, i.e., a stable state with
the order parameter close to that of state (\ref{final state})\ emerges when
the source and drain are balanced. Numerical simulation is performed to
verify our predictions. We compute the time evolution $\left\vert \psi
\left( t\right) \right\rangle =e^{-i\left( H_{\mathrm{T}}+\mathcal{H}_{%
\mathrm{quen}}\right) t}\left\vert \psi \left( 0\right) \right\rangle $ by
exact diagonalization. The geometries of finite systems are schematically
illustrated in Fig. \ref{fig1}. We consider an $N=9$ site system with
periodic and open boundary conditions. In this case, the order parameter has
the following explicit form:%
\begin{equation}
\mathcal{O}(t)=\frac{1}{4}\sum_{n=1}^{4}\frac{|\left\langle \psi \left(
t\right) \right\vert c_{\frac{2\pi n}{9}}c_{-\frac{2\pi n}{9}}\left\vert
\psi \left( t\right) \right\rangle |}{\left\langle \psi \left( t\right)
\right\vert \psi \left( t\right) \rangle }.  \label{Ot}
\end{equation}%
We plot $\mathcal{O}(t)$\ as a function of $t$ and $\varphi $ in Fig. \ref%
{fig2} for the finite size cases schematically illustrated in Fig. \ref{fig1}%
, obtained by numerical simulations. The numerical results agree
with our prediction\textbf{, }that for all cases, $\mathcal{O}(t)$\ tends to
stabilize at $\mathcal{O}(\infty )$ after a sufficiently long time. In
addition, we find that the relaxation process and the final $\mathcal{O}$\
depend on the geometry and $\varphi $. (i) $\mathcal{O}${$(\infty )$}\ is
not sensitive to $N_{0}$ for the ring system, and $\mathcal{O}${$(\infty )$}
can reach the maximum $0.5$. (ii) $\mathcal{O}${$(\infty )$}\ depends on $%
N_{0}$ for the chain system, and $\mathcal{O}${$(\infty )$} can reach the
maximum $0.45$ when the pair source and drain are located at the ends of the
chain. This implies that the balanced edge pair source and drain benefit to
forming a superconducting state. For both boundary conditions, the
converging time depends on the value of $\varphi $, but in two different
ways.

\section{Summary}

\label{sec_summary}

In summary, we have studied several types of toy models with deliberately
engineered pair terms to explore the possibility of realizing nonequilibrium
superconducting state in a nonhomogeneous Kitaev model, which is essentially
a non-Hermitian extension of the Kitaev chain. In the framework of quantum
mechanics, based on the analysis of the exact solution and perturbation
method, we find that the EP dynamics provides a clear picture for the action
of a single pair source or drain. When the two coexist simultaneously, the
dynamics depend on the distance and the relative phase between the pair
creation and annihilation terms. Analytical analysis and numerical
simulation results show that the superconducting state can be dynamically
established at the resonant local pair source and drain. Two spatially
separated pair terms can drive an initial empty state to a stationary state
with a near maximal pair order parameter. It provides an alternative way of
understanding the mechanism of the nonequilibrium superconducting state.

\section*{Acknowledgement}

This work was supported by the National Natural Science Foundation of China
(under Grant No. 11874225).

\end{document}